\documentclass[a4paper,12pt]{article}
\usepackage{authblk}
\usepackage{array}
\usepackage{booktabs}
\usepackage{graphicx}
\begin{document}

\title{{\Large Wood-Saxon alpha potential for p-nuclei $^{106}$Cd and $^{113}$In
}}
\author{\large Dipali Basak}
\author{\large Chinmay Basu}
\affil{Saha Institute of Nuclear Physics, HBNI, 1/AF, Bidhannagar, Kolkata-700064}
\date{}
\maketitle
 {Alpha elastic scattering of p-nuclei were studied for calculating optical potentials. Choice of the $\alpha$-optical potentials are important to measure the reaction rates of p-process. $^{106}$Cd$(\alpha,\alpha)^{106}$Cd and $^{113}$In$(\alpha,\alpha)^{113}$In elastic scattering cross-section data were used to determine the potential parameter sets at E$_{lab}$= 16.1-27 MeV for $^{106}$Cd and  E$_{lab}$=16.14, 19.4 MeV for $^{113}$In system. A Wood-Saxon potential form factor is used for both real and imaginary part. The potential parameters extracted in the present study exhibit a satisfactory result with respect to existing global potential parameters.}


\section{Introduction}

Nuclei heavier than iron(Z$>$26) are produced in the star by neutron capture process, either by slow neutron capture(s-process) or rapid neutron capture(r-process)\cite{a}. But there are 35 naturally occurring isotopes($^{74}$Se-$^{196}$Hg) which are not produced by s or r-process. They are situated in the neutron deficient side of the valley and their natural abundances are small compared with the s and r-nuclides, these nuclides are known as p-nuclei. They are produced in ($\gamma$,p),($\gamma$,$\alpha$),($\gamma$,n) reaction process from s and r-seed nuclides in a high $\gamma$-flux scenario-called photodisntegration process \cite{b}.\\
Reaction rates derived from the reaction cross-sections are necessary inputs for the network calculation of the p-process. The $\gamma$-induced reaction cross-sections can be obtained from the inverse reaction cross-section using principle of detailed balanced. Measurements of ($\gamma$,p),($\gamma$,$\alpha$),($\gamma$,n) reactions on p-nuclei are therefore important to understand the $\gamma$-process. In order to analyse measured cross-sections one has to calculate them using the statistical model for compound nuclear processes. These calculations are sensitive to the entrance channel optical potentials involving $\alpha$ or protons interacting with p-nuclei. The optical potentials for $\alpha$ and proton is very well studied in the literature in terms of generic targets. It is interesting to examine the validity of the global proton and alpha potentials on p-nuclei targets at different energies. In fact there are very few measurements that have extracted the optical potentials for p-nuclei targets. G.G Kiss et al. measured $\alpha$-elastic scattering from $^{106,110,116}$Cd nuclei at E$_{lab}$=16.1,19.6 MeV \cite{c} and on $^{106}$Cd at E$_{lab}$=16.1,17.7,19.6 MeV \cite{d}. A.Ornelas et al. measured elastic scattering of $\alpha$-particles at E$_{lab}$= 16.1-27 MeV on $^{106}$Cd nuclei  \cite{e}. A. Palumbo et al. made a systematic study of $\alpha$-elastic scattering on $^{106}$Cd,$^{118}$Sn,$^{120,124,126,128,130}$Te nuclei at 17-27 MeV incident energies \cite{f}.
G.G Kiss et al. have uses a microscopic model for the real potential viz. DDM3Y for Ref.\cite{d} and M3Y for Ref.\cite{c}
 and Wood-Saxon form factor for imaginary potential. They have only analysed data up to 19.6 MeV and mentioned their $\chi^2$ value. A. Ornelas et al.\cite{e} used DDM3Y potential again but used surface imaginary potential only at low energy, but both surface(W$_s$) and volume imaginary(W$_v$) potential at higher energy. They obtained a low $\chi^2$ at higher energy(23-26 MeV). In the work of Palumbo et al.\cite{f} again the real potential is taken as M3Y with an energy dependence. The imaginary part is parameterized in terms of energy(E$_{\alpha}$) and mass number(A). Five parameters have been used in defining  W$_s$ and W$_v$. G.G Kiss et al. has measured $\alpha$-elastic scattering on $^{113}$In with the same procedure as for Cd and Te targets at 16.14 and 19.4 energies \cite{i}. D. Galaviz et al. studied $\alpha$-elastic scattering on  $^{112,124}$Sn at E$_{\alpha}$=14.4, 19.5 MeV \cite{g} again using the folding model parameterization for real potential. The only work that uses Wood-Saxon form factor is for Sn cases by Avrigeanu et al.
 at 14.4 and 19.5 MeV \cite{h}.\\
 In this work we study the $\alpha$-elastic scattering of $^{106}$Cd and $^{113}$In using a Wood-Saxon form factor for both real and imaginary potentials. This is unlike to all previous calculations who used a microscopic folding model for the real potential. Some of these authors have used energy and mass parameterization in the imaginary potential. A global parameter set is suggested for the system to examine measurements at various energies.

\section{Alpha nucleus optical potential for the present system}

The optical model potential is given by 
\begin{equation}
 U(r)=V_c(r) + V(r) + iW(r) 
 \end{equation}
 V$_c$(r) is the coulomb potential.\\
 V(r) and W(r) are the real and imaginary part of the nuclear potential. For real and imaginary part used Wood-Saxon potential form. 
 \begin{equation}
 V(r)=-V_of_v(r)   
  \end{equation}
 \begin{equation}
 W(r)=(-W_vf_w(r)+W_s\frac{df_w}{dr})  
 \end{equation}
 where 
 \begin{equation}
 f_i(r)=\frac{1}{1+exp\frac{r-R_i}{a_i}} \hspace{1.2 cm}  i = v,w 
 \end{equation}
 Where $R_i$ and $a_i$ are the radii and diffusivities respectively.
 $V_o$ ,$W_v$ , $W_s$ are the potential depths of real,volume imaginary and surface imaginary respectively.\\
  Spin-orbit interaction term included in the optical potentials for spin non zero target/projectile. The form of spin-orbit potential 
  \begin{equation}
   V_{so}(r) = V_{so}{\left(\frac{\hbar}{m_{\pi}c}\right)}^2\frac{1}{r}\frac{df_{so}(r)}{dr} \vec{l}\cdot\vec{s}
   \end{equation}
   \begin{equation}
   {\left(\frac{\hbar}{m_{\pi}c}\right)}^2 \approx 2.00\hspace{0.5 cm}, \hspace{0.5 cm} f_{so}(r) = \frac{1}{1+exp\frac{r-R_{so}}{a_{so}}}
  \end{equation}  
 
 The experimental elastic scattering data of $^{106}$Cd and $^{113}$In fitted using the optical parameter search SFRESCO \cite{sf}.

\section{Results and Discussions}

The results of the SFRESCO fit/calculations \cite{sf} are shown in  Fig. 1 for $\alpha$ + $^{113}$In and Fig. 2(a)-(h) for $\alpha$ + $^{106}$Cd. The results of the calculations performed with well known global potentials of McFadden and Satchler \cite{j} and that of Avrigeanu \cite{k} are shown by dashdotted and dashed lines. The fit/calculations of the present work are shown by solid lines.\\
The $\chi^2$ values for the present calculations are shown in {\it Table 1.} and {\it Table 2.} for $^{113}$In and $^{106}$Cd respectively. It is found that  $\chi^2$ value worsen from 22 to 27 MeV. A reason for this may be due to the nuclear diffraction patterns at backward angles which makes the fit more difficult and worsen in cases compared to when the coulomb dominates at lower energies.\\
The addition of surface imaginary improves the higher energy $\chi^2$ by about 10\%. This addition do not offer any further change in the $\chi^2$ at lower energies.\\
There is however no higher energy data for $^{113}$In where this effect can be studied. The effect of spin orbit potential is not significant for this system.

\begin{figure}[htp]
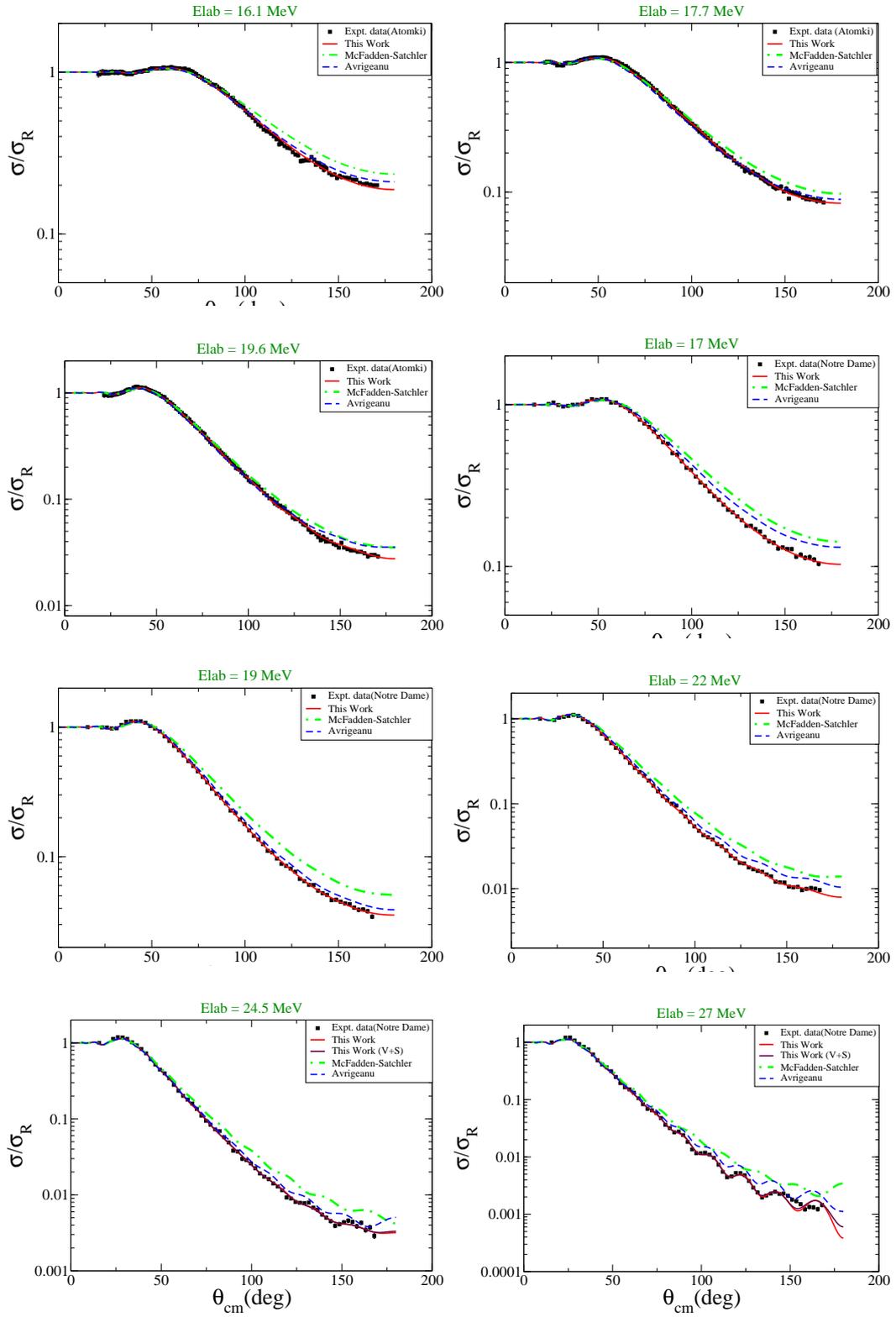

\begin{center}
 \begin{tabular}{cc}
\includegraphics[width=0.50\textwidth]{fig1}
\includegraphics[width=0.50\textwidth]{fig2}\\
\includegraphics[width=0.50\textwidth]{fig3}
\includegraphics[width=0.50\textwidth]{fig4}\\
\includegraphics[width=0.50\textwidth]{fig5}
\includegraphics[width=0.50\textwidth]{fig6}\\
\includegraphics[width=0.50\textwidth]{fig7}
\includegraphics[width=0.50\textwidth]{fig8}\\
\end{tabular}
\caption{$\alpha$ + $^{106}$Cd Elastic scattering data}
\end{center}
\end{figure} 
 
\begin{figure}[htp]
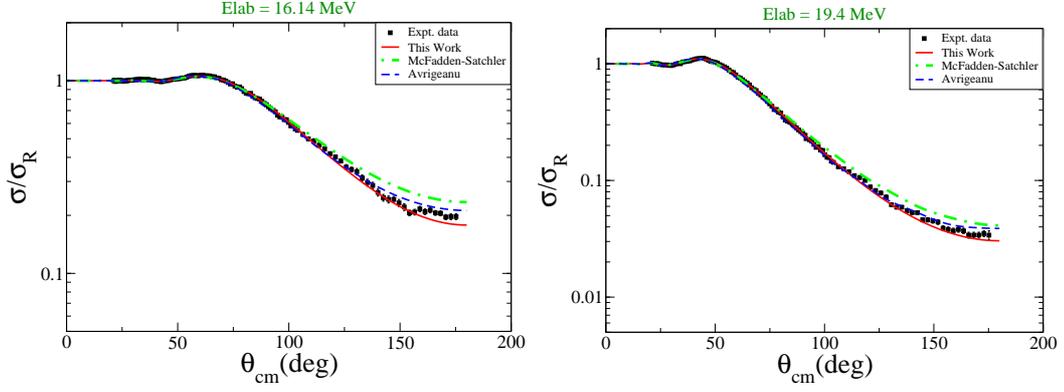

\begin{center}
\begin{tabular}{cc}
\includegraphics[width=0.50\textwidth]{fig9}
\includegraphics[width=0.50\textwidth]{fig10}\\
\end{tabular}
\caption{$\alpha$ + $^{113}$In Elastic scattering data}
\end{center}
\end{figure} 

\begin{table}[!]
  \centering
  \caption{Optical potential parameters of $\alpha$ + $^{106}$Cd system}
  \small
    \begin{tabular}{p{0.06\textwidth}lllllllllp{0.1\textwidth}} 
    \toprule
         \textbf{E$_{lab}$} & \multicolumn{3}{c}{\textbf{Real part}} &\multicolumn{6}{c}{ \textbf{Imaginary part}}
      &\textbf{$\chi^2$/N}\\
     (MeV) & V$_0$ & R$_v$ & a$_v$ & W$_v$ & R$_w$ & a$_w$ &W$_{s}$ & R$_{s}$ & a$_{s}$ &(This work) \\
      
      \midrule
      Atomki\\
      16.1 & 88.01 & 1.07 & 0.60 & 17.76 & 1.16 & 0.45 &&&& 1.698 \\
      17.7 & 73.77 & 1.10 & 0.56 & 10.02 & 1.15 & 0.46 &&&&  1.122\\
      19.6 & 68.94 & 1.10 & 0.57 & 10.81 & 1.17 & 0.45 &&&&  1.098\\
      NotreDame \\
      17.0 & 85.91 & 1.10 & 0.60 & 23.87 & 1.18 & 0.45 &&&&   0.998 \\
      17.65 & 88.77 & 1.09 & 0.60 & 30.00 & 1.15 & 0.45 &&&&  1.076 \\
      19.0 & 71.21 & 1.11 & 0.60 & 13.21 & 1.21 & 0.45 &&&&    1.092\\
      22.0 & 67.91 & 1.12 & 0.60 & 12.41 & 1.21 & 0.46 &&&&   2.065 \\
      24.5 & 63.55 & 1.10 & 0.60 & 12.88 & 1.18 & 0.45 &&&& 4.370\\
      27.0 & 68.85 & 1.07 & 0.62 & 11.95 & 1.16 & 0.55 &&&&  12.402\\
      \midrule

       24.5 & 83.09 & 1.10 & 0.57 & 14.64 & 1.09 & 0.40 & 6.41 & 1.14 & 0.38 &  4.153\\
      27.0 & 61.57 & 1.13 & 0.57 & 8.68 & 1.11 & 0.53 & 11.52 & 1.12 & 0.38 &  10.960\\
            \bottomrule
    \end{tabular}
\end{table}

\begin{table}[h]
  \centering
  \caption{Optical potential parameters of $\alpha$ + $^{113}$In system}
  \small
    \begin{tabular}{p{0.06\textwidth}lllllllllp{0.1\textwidth}} 
    \toprule
      \textbf{E$_{lab}$} & \multicolumn{3}{c}{\textbf{Real part}} &\multicolumn{3}{c}{ \textbf{Imaginary part}} & \multicolumn{3}{c}{ \textbf{Spin-Orbit term}}
       &\textbf{$\chi^2$/N}\\
    (MeV)  & V$_0$ & R$_v$ & a$_v$ & W$_v$ & R$_w$ & a$_w$ &V$_{so}$ & R$_{so}$ & a$_{so}$  &(This work) \\
      
      \midrule
      16.14 & 78.38 & 1.10 & 0.57 & 11.33 & 1.11 & 0.39  & 5.004 & 1.029& 0.64 &  1.163 \\
      19.4 & 118.59 & 1.09 & 0.54 & 24.55 & 1.15 & 0.36 & 5.19 & 1.048 & 0.48 &  1.460\\
            \bottomrule
    \end{tabular}
\end{table}

\paragraph{}
A global potential parameter sets is suggested for $^{106}$Cd system.
 $$V_0 \approx 320.88 -22.58E_{cm} +0.4958{E_{cm}}^2$$ 
$$R_v \approx 1.10 \hspace{0.3 cm},\hspace{0.3 cm} a_v \approx 0.60 $$
$$W_v \approx 111.168 -8.6602E_{cm} +0.18763{E_{cm}}^2$$ 
 $$R_w \approx 1.18 \hspace{0.3 cm},\hspace{0.3 cm} a_w \approx 0.45 $$
\begin{figure}[h!]
\begin{center}
   \includegraphics[width=0.50\textwidth]{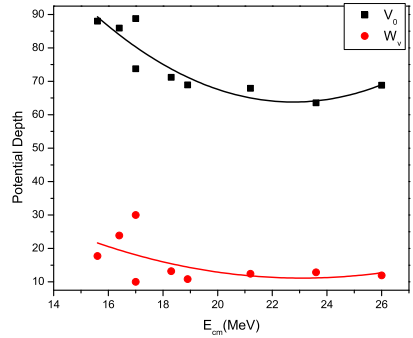}
    \caption{Variation of V$_0$ and W$_v$ with CM energy}
    \end{center}
 \end{figure}
       
The real(V$_0$) and imaginary(W$_v$) potential depths at different CM energies are fitted to obtain a energy dependence global form for the two potentials.
 As the values of R$_i$ and a$_i$ are almost constant with energy variation their average values are used in the global parameter set. A global potential for $^{113}$In can be revisited with more measurements at higher energies.

\end{document}